# Navigating in the Dark – Designing Autonomous Driving Features to Assist Old Adults with Visual Impairments

Lashawnda Bynum, Jay Parker, Kristy Lee, Nia Nitschke, Melanie LaFlam, Jennifer Marcussen, Jana Taleb, Aleyna Dogan, Lisa J. Molnar, Feng Zhou

Age-related macular degeneration is a leading cause of blindness worldwide and is one of many limitations to independent driving among old adults. Highly autonomous vehicles present a prospective solution for those who are no longer capable of driving due to low vision. However, accessibility issues must be addressed to create a safe and pleasant experience for this group of users so that it allows them to maintain an appropriate level of situational awareness and a sense of control during driving. In this study, we made use of a human-centered design process consisting of five stages - empathize, define, ideate, prototype, and test. We designed a prototype to aid old adults with age-related macular degeneration to travel with a necessary level of situational awareness and remain in control while riding in a highly or fully autonomous vehicle. The final design prototype includes a voice-activated navigation system with three levels of details to bolster situational awareness, a 360˚ in-vehicle camera to detect both the passenger and objects around the vehicle, a retractable microphone for the passenger to be easily registered in the vehicle while speaking, and a physical button on the console-side of the right and left front seats to manually activate the navigation system.

## INTRODUCTION

Visual impairment is increasingly prevalent among older adults due to age-related macular degeneration (AMD) that affects the central retina. AMD is a leading cause of blindness worldwide (Lim et al., 2012) and occurs in 1.5% of the population in the United States over the age of 40, but has increased to more than 15% in certain demographic groups, such as white women over the age of 80 (Friedman et al., 2004). Visual impairment limits the ability of older adults to perform tasks in everyday life, such as reading, recognizing faces, and driving (Owsley & McGwin, 2008). A previous study surveying a subset of the AMD population found a strong association between severity of AMD and self-rated difficulty in driving (Mangione et al., 1999). Visual impairment in old adults was also found to be associated with a higher rate of motor vehicle crashes (Swain et al., 2021).

However, it is often difficult for adults with visual impairment to give up driving completely despite the issues they face while operating a vehicle. Kim (2011) found that only a third of old adults might restrict their activities due to lack of transportation options [5]. Moreover, driving cessation in old adults might contribute to health-related issues, such as depression (Chihuri et al., 2016). With increasingly more old adults with possible visual impairment, there is an urgent need for adequate resources to help meet their transportation needs.

Currently, conventional vehicles do not provide solutions for old adults with visual impairments. Autonomous vehicles might provide a transportation solution to improve the mobility of old adults with different levels of visual impairment and other disabilities (Padmanaban et al., 2021). Nevertheless, many vehicles on the road are still below SAE (Society of Automotive Engineers) Level 3 and have automation features that still require drivers to remain attentive during driving (Ayoub et al., 2019). Even with SAE Level 3 vehicles, the driver is still required to take over control whenever requested by the vehicle and this poses a great challenge (Ayoub et al., 2022), especially when the driver is out of the control loop without enough situational awareness (Avetisyan et al., 2022). Maintaining situational awareness is especially difficult for old adults with visual impairments. For highly autonomous vehicles in certain geo-fenced areas (i.e., SAE Level 4), people still find it difficult to trust the vehicle (Ayoub et al., 2021; Zhang et al., 2022), especially older adults (Molnar et al., 2017).

The U.S. Department of Transportation promotes innovative design solutions to help people with disabilities, such as visual impairment to improve their mobility, especially for highly autonomous vehicles (Padmanaban et al., 2021). Thus, in this paper, we attempted to design a prototype of highly autonomous vehicles (SAE Level 4) through a human-centered design process in order to promote trust and acceptance of old adults with visual impairment. In order to understand how highly autonomous vehicles can be used to improve mobility of old adults with visual impairment, the objectives of this study are summarized as follows:
1) Explore and understand behavioral patterns, user needs, and pain points of old adults with visual impairment in a conventional vehicle, and
2) Design, develop, and evaluate a prototype of a highly automated vehicle to help solve the most pressing barriers for old adults with visual impairment.

## METHOD AND RESULTS

In this study, we followed a human-centered design process from Stanford University's d.school, i.e., empathize, define, ideate, prototype, and test (Padmanaban et al., 2021). These five steps comprised the underlying process to create our design solutions. We included multiple iterations on some steps to clarify and refine our design challenge and prototype.

**Empathize**

To better understand the experience of old adults with visual impairment while driving, we targeted people with visual impairment who were 65 years of age or older for our first round of interviews. Our goal was to understand how this group of users navigated driving tasks in response to their changes in vision. We recruited five individuals for our interviews. Four participants were in the U.S. and one participant was based in South Korea. Each had some level of visual impairment which made driving more difficult ($n = 4$) or had caused them to stop driving altogether ($n = 1$).

In the interview, we aimed to understand what role vehicles played in their lives, what options they had for transportation, and how they had adjusted their driving behaviors to accommodate their changes in vision. We also examined the participants' experiences in using existing advanced driver-assistance systems (ADAS) and their opinions on autonomous vehicles in general. Table 1 summarizes the findings from the interviews.

**Table 1. Summary from the interviews**

| Topic | Summary of Responses |
|---|---|
| Role of vehicles in daily life | Sense of freedom and control; Leading active life and riding with others when possible, but still fully rely on cars to meet daily needs |
| Issues driving vehicles | Visual impairment; Distracted driving when making changes in vehicle settings or devices; Accessibility; Increased reliance on auditory cues to compensate for visual loss |
| Adjustments to driving as vision changed | Changed driving behavior (avoid driving at night, taking known routes only); Relying on visual aids (glasses or bifocals); Physically adjusting, squinting, or changing positions in vehicle to better see surroundings |
| If still driving, what would make you stop? | Cognitive and physical decline; Visual decline; Societal pressure to stop; Other forms of transportation becoming accessible |
| Experience with voice controls | Distracting and inaccurate; Reliance on external applications/devices instead of in-vehicle ones |
| Experience with ADAS | Avoidance to gain a sense of control; Safety concerns; Could be useful if designed for specific user types |
| Learning about new technology | Avoidance; Hands-on learning or in-person instruction (dealerships); Self-service/internet/manuals; Auditory learning & feedback |

All the participants considered vehicles essential in their life and expressed that they were still leading active lifestyles outside home, needing to travel frequently, sometimes long distances, to see family, to attend gatherings and events, or to work. A common theme was the desire for "freedom" brought by vehicle mobility and "control of the vehicle" while in the vehicle. The desire for freedom seemed especially important for U.S.-based participants who noted that without a vehicle, they would not be able to go anywhere ("*Without my car, I cannot go to my doctors, grocery stores, drug stores, or visit my family or friends*"). The desire for a sense of control indicated that the participants trusted their driving abilities and their doubts about automation features, especially when the participants were asked about their perception of ADAS ("*I'm a good driver. I like being in control...Turn a lot [of the ADAS features] off because they irritate me*").

All the participants recognized the issues while driving with different levels of vision loss. Four of the participants were still able to drive during the day, but complained that it was difficult or distracting for them to adjust settings in the vehicle and that they had to rely on auditory cues for directions while driving in unfamiliar surroundings. They also reported using some form of corrective lenses to supplement their vision (corrective lenses, bifocals) as part of adjusting their driving behaviors. One participant noted, "*I use my corrective lenses which bring my vision to its best self, but still found it difficult to drive during the night.*" Behavioral adjustments (e.g., physically changing position, squinting) were also made in addition to wearing their glasses: "*I reduce my speed at night so I don't come too close to an object especially on my left side*". Although trying their best to keep driving, these four participants reported they would stop driving, when there was further visual decline, other cognitive and physical decline, and family pressures. One participant had to give up driving due to visual impairment and hoped to increase accessibility by using different types of transportation.

We also aimed to understand their experience with automation features in the vehicle and their attitudes toward autonomous vehicles. The participants worried about the safety concerns of ADAS and would avoid using them to gain a sense of control. They also would have to rely on external voice controls to reduce inaccuracy ("*I would really need stuff in the car, [and I would tell voice control] like tell me how to get somewhere, or call somebody, and I feel like it never got what I wanted. So I was like okay this is frustrating, I'm going to turn this off*"). They also had a low level of trust in autonomous vehicles ("*[I] wouldn't own an autonomous vehicle, ever!...Never been in one…*", "*I would be hesitant to be in such a vehicle, unless it proves itself to be safe and I can still be in control*"). The participants felt they were more in control of the vehicle than the technologies that currently or even in the future to assist them.

When asked about how they attempt to learn to work the vehicle features, most participants were hands-on learners, preferring to use a feature and learn how it worked by doing: "*I sit down and read some of it and then get in the car and demonstrate it myself*". While there was initial negative feedback about voice control in the vehicle, many participants stated auditory cues would help them learn, similar to how GPS navigation apps work, such as Google Maps: "*The way I know that I've arrived at the destination is by navigation sound. When it tells me that I've arrived, I know I'm there*".

**Define**

From the interviews, it was clear that freedom and control were important to old adults when it came to transportation. We thus (re)framed our design challenge to focus on creating a sense of freedom and control for individuals with declining visual impairment that might not allow them to drive anymore. To more clearly define and empathize with our target audience, we created a persona, Kelly Smith, who was forced to give up driving after 40 years when she was diagnosed with age-related macular degeneration. Kelly wished she could resurrect the joys and ease of driving by herself, as she at times felt like a burden to her family who had to drive her around.

With Kelly in mind, we focused on highly (i.e., SAE Level 4) autonomous vehicles since Kelly was no longer able to drive. In such a vehicle, Kelly does not drive at any point and the vehicle can drive autonomously under all conditions in geo-fenced areas. This meant that our design solution should help create a sense of control in the vehicle where the user had no need for a steering wheel, gas pedal, or brake pedal.

Using the persona, we created a scenario that Kelly typically encountered while in an autonomous vehicle to aid our idea generation. The scenario focused on start-up and navigation features in the vehicle, as these were commonly mentioned areas in our initial interviews.

1) Kelly is ready to visit her new friend. She is able to find and get into her vehicle with the help of her guide dog.
2) Kelly gets into her vehicle and would like to know that the vehicle is aware that she and her guide dog are inside.
3) Kelly wants to communicate with the vehicle her intended destination and which route she would like to choose. She does not feel comfortable going on the highway on her way, so she chooses a local route.
4) She also wants to make sure everything is in order before the vehicle drives off. She's worried her vehicle might not have enough battery/fuel to complete the trip and would appreciate reassurance from the vehicle.
5) As the vehicle is driving, Kelly realizes she's a little anxious because she is not familiar with the route and would like details of where she is and every step of the way.

**Ideate**

With this scenario in mind, we brainstormed design solutions using the Crazy Eights method. For a total of eight minutes, we each generated eight designs averaging about a minute per design. From our designs, we each chose our favorites to present to the group as possible solutions to our design challenge. The possible design solutions included:

1) Train guide dogs to control buttons in the vehicle.
2) 24/7 hotline to assist drivers with questions.
3) Vibrating seats to tell users when the vehicle is going to turn.
4) Button on dashboard to turn on navigation.
5) Detailed voice navigation system.
6) 3D model of terrain outside of the vehicle that users can "feel."
7) Heat map on dashboard showing traffic around the vehicle.
8) Checklist of vehicle settings after start-up.
9) Showing a zoomed in map of the vehicle's full route on the windshield.
10) Flashing lights to communicate with the driver.
11) Use of different sound effects to communicate with the driver.

After each team member was given a chance to present their ideas, we collectively voted for what we wanted to implement in the prototype. The ideas that had the most votes were: 1) button on dashboard to turn on navigation (in case the vehicle did not turn on with voice activation), 2) checklist of vehicle settings, and 3) detailed voice navigation system.

**Prototype and Test**

Table 2. Example scenarios and voice prompts

| Scenarios | Voice Prompt |
|---|---|
| Scenario 1: This is your first time in front of a wheel since you stopped driving 3 years ago. This new vehicle is designed to be fully controlled by voice commands. How would you begin to turn the vehicle on? | Welcome to the Mazda CX-60! In full control mode, I, Jay, drive for you. I am your personal driver, you tell me where you'd like to go and I take you there. |
| Scenario 2: The vehicle turns on and is now going to run through a checklist asking you various questions before you take off. Play vehicle voice commands for the participant. | Camera is on and I'm scanning your surroundings. I see that you are in the vehicle. Let's run through a preparation checklist before you take off. Item 1 of 3: Battery is at 100% so you're good to take off. Item 2 of 3: We are putting on your seatbelt, are you comfortable? [user responds] Item 3 of 3: Would you like to turn on the radio and A/C? [user responds] Checklist is all done. Now let's get your navigation level set up. Before the vehicle makes any movement, I will explain its actions. There are two levels of detail. Choose the one you're most comfortable with. [user responds] |

We followed an iterative process of testing our prototypes to learn from our users with a combination of a semi-structured user interview and a Wizard of Oz method. In the Wizard of Oz method, the participants interacted with a system they believed to be autonomous but in reality was controlled by a human operator (Ayoub et al., 2020). We designed five scenarios and voice prompts to best suit our participants while interacting with an autonomous vehicle. We read aloud the scenarios and played the pre-recorded voice prompts. In order to test our prototype, we designed five scenarios and each presented a task for the participant to complete, including 1) prompt to turn vehicle on with voice command, 2) vehicle checklist: vehicle diagnostics, seatbelt, cabin conditioning (AC, radio, seat adjust), and prompt navigation level of details select: basic or detailed, 3) prompt to provide destination and select a route and prompt to accept or cycle through additional route options, 4) inform user of trip start and vehicle action (i.e. backing out of parking space), 4) inform user of navigational movements; prompt user for any actions to take en route (i.e. turn on red, take detour). We also showed example scenarios and voice prompts in Table 2 associated with these tasks.

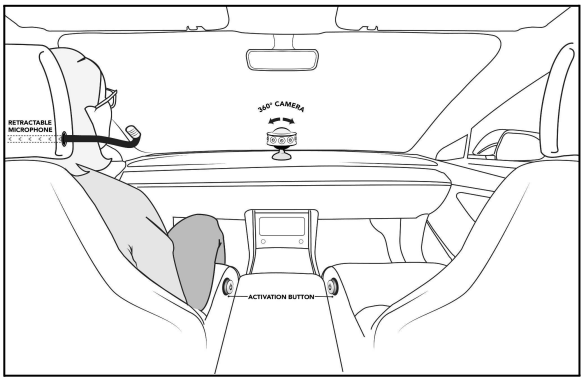

Figure 1. Sketch of the vehicle's interior of the initial prototype

We recruited four participants over 65 years old with some level of visual impairment to test our low-fidelity prototype as sketched in Figure 1, which included a voice-activated navigation system with two different levels of details, an in-vehicle camera to monitor the situation in the vehicle, a retractable microphone for the user to be easily registered and control the vehicle, and a physical button on the console-side of the right and left front seats to manually activate the autonomous driving system.

We described the overall configuration of the vehicle's interior as shown in Figure 1. The participants liked the checklist and possible details of navigation during the driving. They also liked the position of the button on the side of the seat because it was easy to reach (see Figure 1). One participant previously had a stroke, leading to physical impairment in the left side of his body. He pointed out that it would be better to have buttons on both seats to create a choice for participants to sit on the side that best aided their mobility difficulties.

Then, we placed the participants in different scenarios for them to go through all the tasks with the Wizard of Oz method. All the participants appreciated how prepared they felt before the vehicle took off due to the vehicle's initial checklist. One noted, "*The checklist at the beginning was super comfortable. There wasn't anything that was unnecessary*", while another said that, "*The initial checklist was good, especially the detail to start up the vehicle*". The level of voice detail also reassured participants: "*I like the narration as you go. If you don't hear it, you don't know what the car knows. I thought the reassurance was a good thing*".

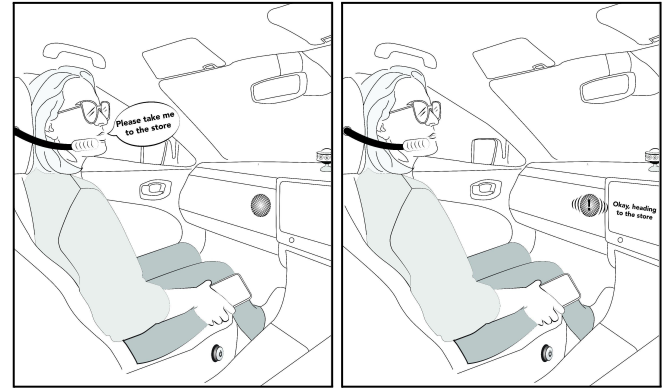

(a)                          (b)

Figure 2. Sketch of user interacting with the vehicle using voice commands: (a) A speech command by the user, (b) The response from the vehicle

Although participants felt aware of their surroundings before heading off for a ride, two participants worried about a lack of situational awareness while in motion. One participant worried about being unable to control the vehicle while moving, "*Can you tell the car to slow down if it was going too fast?*" Another participant wanted to make sure the vehicle could distinguish between an inanimate object and a pedestrian: "*At this time, the sensors for autonomous vehicles cannot distinguish between humans or objects…hopefully it will be resolved in the near future*". Finally, while two participants appreciated the level of details in the voice navigation system, others did not. One noted that, "*I thought it was too verbose…you just want to get to where you're going*".

Based on the feedback from the participants, we refined our prototype. First, we added in a sensor system to alert the passenger to when pedestrians or objects were in front of the vehicle to address concerns about awareness while the vehicle was in motion (see Figure 3). Second, we created an additional level of navigation that included fewer details than either of the previous two levels (as shown in Table 3). Third, we changed the naming convention of the navigation levels to take the cognitive burden off of passengers in remembering what each level represents. The refined prototypes were then presented to the participants with revised tasks and their satisfaction level was much improved.

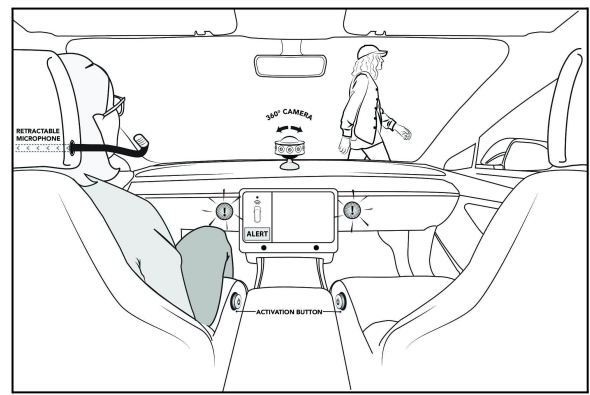

Figure 3. Sketch of enhanced sensor system in vehicle's interior

Table 3. Changes made to levels of detail after testing

| Scenarios | Voice Prompt |
| --- | --- |
| Limit | Information about route changes |
| Basic | Hazards, weather conditions, speed limits, some information about vehicle behavior |
| Detailed | Road types, speed limits, hard braking & acceleration, weather conditions, oncoming emergency vehicles, police officers, hazards/construction, detailed information about vehicle behavior. |

## DISCUSSIONS AND CONCLUSIONS

In this study, we aimed to design a solution to help bolster the feeling of control and a sense of freedom for old drivers with visual impairment to feel more comfortable navigating within autonomous vehicles. Through a human-centered design process, we were able to gain insight into the needs and preferences of our target user group and iteratively improve our design to better meet their needs.

Through our empathy stage, we identified the major behavioral patterns, pain points, and needs of old adults with visual impairment with regard to driving. Even though autonomous vehicles hold promise for improving their mobility and independence, old adults were reluctant to trust them without a sense of control and understanding of the overall situational awareness during driving. Based on such findings, we defined our design challenge to generate corresponding ideas with Crazy Eights. We came up with voice prompts to provide situational awareness during driving to improve the sense of control and freedom. Through testing such an idea using a Wizard of Oz method, we found that extra details did not necessarily increase situational awareness, but simple, precise information did. Users expressed interest in the vehicle system providing feedback primarily about important changes in the environment or traffic events they could choose to react to. They were less responsive to the vehicle providing constant narration of its navigational actions. We also found that the participants would have a better sense of control by customizing their ride experience based on their preferences and thus increase their comfort and trust in the vehicle, such as the ability to recall their presetting to reduce the time it took to engage the system and begin a trip, the inclusion of a sensor to alert the passenger about the objects and pedestrians around the vehicle, and the options to take a preferred route.

However, our research was not without limitations. Due to time constraints, we were unable to test our prototype with a larger and more diverse pool of participants. Additionally, not all of our participants had a level of visual impairment that prevented them from driving in the empathy and testing stages. In the testing stage, we somehow covered their eyes to better simulate the scenarios using the Wizard of Oz method. Furthermore, the testing scenarios we used involved minimal driving, so it was unclear how well our prototype would perform in more complex real-world situations. Other limitations include the lack of highly autonomous vehicles, as they are not currently available for consumer purchase.

These limitations highlight the need for further research in this area to focus more on testing the effectiveness of our design prototype in a larger and more diverse sample of participants in more scenarios. Additional research could also be conducted to explore other potential design solutions for old adults with visual impairment. Ultimately, our goal is to create accessible and user-friendly technologies that can help improve the lives of old adults with age-related macular degeneration and other visual impairments, and enable them to maintain their independence and mobility.